\documentclass[prb,showpacs,preprintnumbers,amsfonts,amssymb,amsmath,floats,twocolumn,aps]{revtex4}

\usepackage{graphicx}
\usepackage{dcolumn}
\usepackage{bm}
\usepackage[final,dvips]{epsfig}
\usepackage{bm}
\usepackage{color}

\newcommand{\be}{\begin{equation}}
\newcommand{\ee}{\end{equation}}
\newcommand{\ba}{\begin{eqnarray}}
\newcommand{\ea}{\end{eqnarray}}
\newcommand{\ff}[1]{{\bm #1}}
\newcommand{\tr}{\mbox{tr}}
\newcommand{\Tr}{\mbox{Tr}}
\newcommand{\refeq}[1]{Eq.\ (\ref{eq:#1})}

\begin{document} 
  
\title{Non-perturbative conserving approximations and Luttinger's sum rule} 

\author{Jutta Ortloff, Matthias Balzer, Michael Potthoff}

\affiliation{
Institut f\"ur Theoretische Physik und Astrophysik, 
Universit\"at W\"urzburg, Am Hubland, D-97074 W\"urzburg, Germany
}
 
\begin{abstract}
Weak-coupling conserving approximations can be constructed by truncations of the 
Luttinger-Ward functional and are well known as thermodynamically consistent approaches
which respect macroscopic conservation laws as well as certain sum rules at zero temperature.
These properties can also be shown for variational approximations that are generated
within the framework of the self-energy-functional theory without a truncation of the 
diagram series.
Luttinger's sum rule represents an exception.
We analyze the conditions under which the sum rule holds within a non-perturbative
conserving approximation. 
Numerical examples are given for a simple but non-trivial dynamical two-site approximation. 
The validity of the sum rule for finite Hubbard clusters and the consequences for cluster 
extensions of the dynamical mean-field theory are discussed.
\end{abstract} 
  

\maketitle 

\section{Introduction}
\label{sec:intro}

Continuous symmetries of a Hamiltonian imply the existence of conserved quantities:
The conservation of total energy, momentum, angular momentum, spin and particle number
is enforced by a not explicitly time-dependent Hamiltonian which is spatially homogeneous 
and isotropic and invariant under global SU(2) and U(1) gauge transformations.
For the treatment of a macroscopically large quantum system of interacting fermions, 
approximations are inevitable in general. 
Approximations, however, may artificially break symmetries and thus lead to unphysical 
violations of conservations laws.

Baym and Kadanoff \cite{BK61,Bay62} have analyzed under which circumstances an approximation 
for time-dependent correlation functions, and for one- and two-particle Green's functions 
in particular, respect the mentioned macroscopic conservation laws.
They were able to give corresponding rules for a proper construction of approximations,
namely criteria for selecting suitable classes of diagrams, within diagrammatic 
weak-coupling perturbation theory. 
Weak-coupling approximations following these rules and thus respecting conservation laws
are called ``conserving''.
Frequently cited examples for conserving approximations are the Hartree-Fock or the 
fluctuation-exchange approximation. \cite{BK61,BSW89,BW91}

Baym \cite{Bay62} has condensed the method of constructing conserving approximations
into a compact form:
A conserving approximation for the one-particle Green's function $\ff G$ is obtained
by using Dyson's equation $\ff G = 1 / (\ff G_0^{-1} - \ff \Sigma)$ with (the free,
$\ff U=0$, Green's function $\ff G_0$ and) a self-energy 
$\ff \Sigma = \ff \Sigma_{\ff U}[\ff G]$ given by a universal functional.
Apart from $\ff G$, the universal functional $\ff \Sigma_{\ff U}$ must depend on the 
interaction parameters $\ff U$ only.
Furthermore, the functional must satisfy a vanishing-curl condition or, alternatively,
must be derivable from some (universal) functional $\Phi_{\ff U}[\ff G]$ as 
$T \ff \Sigma_{\ff U}[\ff G] = \delta \Phi_{\ff U}[\ff G] / \delta \ff G$ 
(the temperature $T$ is introduced for convenience).
In short, ``$\Phi$-derivable'' approximations are conserving.

$\Phi$-derivable approximations have been shown \cite{Bay62} to exhibit
several further advantageous properties in addition. 
One of these concerns the question of thermodynamical consistency.
There are different ways to determine the grand potential of the system from the Green's 
function which do not necessarily yield the same result when using approximate quantities.
On the one hand, $\Omega$ may be calculated by integration of expectation values, accessible
by $\ff G$, with respect to certain model parameters.
For example, $\Omega$ may be calculated by integration of the average particle number, 
as obtained from the trace of $\ff G$, with respect to the chemical potential $\mu$.
On the other hand, $\Omega$ can be obtained as 
$\Omega = \Phi + \Tr \ln \ff G - \Tr (\ff \Sigma \ff G)$
without integration.
A $\Phi$-derivable approximation consistently gives the same result for $\Omega$ in both
ways.

At zero temperature $T=0$ there is another non-trivial theorem which is satisfied by
any $\Phi$-derivable approximation, namely Luttinger's sum rule. \cite{LW60,Lut60}
This states that the volume in reciprocal space that is enclosed by the Fermi surface
is equal to the average particle number.
The original proof of the sum rule by Luttinger and Ward \cite{LW60} is based on the
existence of $\Phi$ in the exact theory and is straightforwardly transferred to the
case of a $\Phi$-derivable approximation. 
This also implies that other Fermi-liquid properties, such as the linear trend of the
specific heat at low $T$ and Fermi-liquid expressions for the $T=0$ charge and the spin 
susceptibility are respected by a $\Phi$-derivable approximation.

\begin{figure}[b]
  \includegraphics[width=0.8\columnwidth]{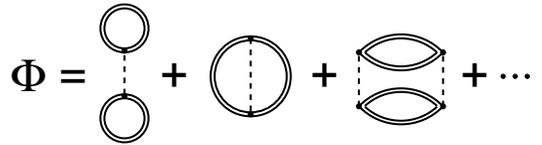}
\caption{
Diagrammatic representation of the Luttinger-Ward functional $\Phi_{\ff U}[\ff G]$.
Double lines stand for the interacting one-particle Green's function $\ff G$, dashed
lines represent the vertices $\ff U$.
}
\label{fig:phi}
\end{figure}

There is a perturbation expansion \cite{LW60,AGD64} which gives the Luttinger-Ward
functional 
$\Phi_{\ff U}[\ff G]$ in terms of closed skeleton diagrams (see Fig.\ \ref{fig:phi}).
As a manageable $\Phi$-derivable approximation must specify a (universal) functional 
$\Phi_{\ff U}[\ff G]$ that can be evaluated in practice, one usually considers truncations
of the expansion and sums up a certain subclass of skeleton diagrams only.
This, however, means that the construction of conserving approximations is restricted
to the weak-coupling limit. 

One purpose of the present paper is to show that it is possible to construct 
$\Phi$-derivable approximations for lattice models of correlated fermions with local 
interactions which are {\em non-perturbative}, i.e.\ do not employ truncations of the 
skeleton-diagram expansion.
The idea is to employ the self-energy-functional theory (SFT). \cite{Pot03a,Pot03b,PAD03} 
The SFT constructs the Luttinger-Ward functional $\Phi_{\ff U}[\ff G]$, or its Legendre transform
$F_{\ff U}[\ff \Sigma]$, in an indirect way, namely by making contact with an exactly
solvable reference system. 
Thereby, the exact functional dependence of $F_{\ff U}[\ff \Sigma]$ becomes available 
on a certain subspace of self-energies which is spanned by the self-energies generated
by the reference system.

The obvious question is whether those non-perturbative $\Phi$-derivable approximations
have the same properties as the weak-coupling $\Phi$-derivable approximations suggested
by Baym and Kadanoff.
This requires the discussion of the following points:

(i) {\em Macroscopic conservations laws}. 
For fermionic lattice models, conservation of energy, particle number and spin have
to be considered.
Besides the static thermodynamics, the SFT concept concentrates on the {\em one-particle}
excitations. 
For the approximate one-particle Green's function, however, it is actually simple to 
prove that the above conservation laws are respected.
A short discussion is given in Appendix \ref{sec:cons}.

(ii) {\em Thermodynamical consistency}.
This issue has already been addressed in Ref.\ \onlinecite{AAPH06a}. 
It has been shown that the $\mu$ derivative of the (approximate) SFT grand potential
(including a minus sign) equals the average particle number $\langle N \rangle$ as 
obtained by the trace of the (approximate) Green's function.
The same holds for {\em any} one-particle quantity coupling linearly via a parameter
to the Hamiltonian, e.g.\ for the average total spin $\langle \ff S \rangle$ coupling 
via a field of strength $\ff B$.

(iii) {\em Luttinger sum rule}.
This is the main point to be discussed in the present paper.
There are different open questions:
First, it is straightforward to prove that weak-coupling $\Phi$-derivable approximations 
respect the sum rule as one can directly take over the proof for the exact theory.
For approximations constructed within the SFT, a different proof has to be given.
Second, it turns out that a non-perturbative $\Phi$-derivable approximation respects 
the sum rule if and only if the sum rule holds for the reference system that is used 
within the SFT.
As the original and thereby the related reference system may be studied in the 
strong-coupling regime, this raises the question which reference system does respect 
the sum rule, i.e.\ which approximation is consistent with the sum rule.
Third, it will be particularly interesting to study reference systems which generate 
dynamical impurity approximations (DIA) \cite{Pot03a,Pot03b} and variational cluster 
approximations (VCA), \cite{PAD03,DAH+04} as these consist of a {\em finite} number 
of degrees of freedom.
Does the Luttinger sum rule hold for finite systems? Do the DIA and the VCA respect 
the sum rule? What is the simplest approximation consistent with the sum rule?
Note that finite reference systems consisting of a few sites only have been shown 
\cite{Pot03b,Poz04,KMOH04,SLMT05,IKSK05a,IKSK05b,AA05,EKPV07} to generate approximations 
which qualitatively capture the main physics correctly.
Finally, it is important to understand these issues in order to understand whether 
and how a violation of the sum rule is possible within cluster extensions 
\cite{HTZ+98,KSPB01,LK00,OMMF03} of the dynamical mean-field theory (DMFT). 
\cite{MV89,GK92a,Jar92,GKKR96,KV04}
Note that the SFT comprises the DMFT and certain \cite{PB07} cluster extensions and 
that possible violations of the sum rule in the two-dimensional lattice models have 
been reported, \cite{PLS98,GEH00,HRKW02} including a study using the dynamical cluster 
approximation (DCA). \cite{MPJ02}

The paper is organized as follows:
A brief general discussion of the Luttinger sum rule is given in the next section, 
and a form of the sum rule specific to systems with a finite number of spatial degrees 
of freedom is derived.
Sec.\ \ref{sec:sft} clarifies the status of the sum rule with respect to non-perturbative 
approximations generated within the SFT framework. 
The results are elucidated by several numerical examples obtained for the most simple but
non-trivial non-perturbative conserving approximation in Sec.\ \ref{sec:two}.
Violations of the sum rule in finite systems and their consequences are discussed in 
Sec.\ \ref{sec:vio}. Finally, Sec.\ \ref{sec:con} summarizes our main conclusions.

\section{Luttinger sum rule}
\label{sec:lsr}

A system of interacting electrons on a lattice is generally described by a Hamiltonian 
$H(\ff t,\ff U)=H_0(\ff t)+H_1(\ff U)$ 
consisting of a one-particle part $H_0(\ff t)$ and an interaction $H_1(\ff U)$ with 
one-particle and interaction parameters $\ff t$ and $\ff U$, respectively.
As a prototype, let us consider the single-band Hubbard model \cite{Hub63,Gut63,Kan63} 
on a translationally invariant $D$ dimensional lattice consisting of $L$ sites with 
periodic boundary conditions. 
The Hamiltonian is given by:
\begin{equation}
  H 
  = 
  \sum_{ij\sigma} t_{ij} c_{i\sigma}^\dagger c_{j\sigma} 
  + 
  \frac{U}{2} \sum_{i\sigma} n_{i\sigma} n_{i-\sigma} \: .
\label{eq:hub}  
\end{equation}
Here, $i=1,...,L$ refers to the sites, $\sigma=\uparrow, \downarrow$ is the spin projection,
$c_{i\sigma}$ ($c_{i\sigma}^\dagger$) annihilates (creates) an electron in the one-electron
state $| i \sigma \rangle$, and $n_{i\sigma} = c_{i\sigma}^\dagger c_{i\sigma}$.
Fourier transformation diagonalizes the hopping matrix $\ff t$ and yields the dispersion 
$\varepsilon(\ff k)$.
There are $L$ allowed $\ff k$ points in the first Brillouin zone.

Let $\ff G = \ff G_{\ff t, \ff U}$ denote the one-electron Green's function of the model 
$H(\ff t,\ff U)$. 
In case of the Hubbard model, its elements are given by
$G_{ij}(\omega) = \langle \langle c_{i\sigma} ; c_{j\sigma}^\dagger \rangle \rangle_{\omega}$. 
In the absence of spontaneous symmetry breaking, the Green's function is spin-independent and
diagonal in reciprocal space.
It can be written as 
$G_{\ff k}(\omega) = 1/(\omega + \mu - \varepsilon(\ff k) - \Sigma_{\ff k}(\omega))$
where $\mu$ is the chemical potential and $\Sigma_{\ff k}(\omega)$ the self-energy. 
We also introduce the notation $\ff \Sigma_{\ff t, \ff U}$ for the self-energy, and
$\ff G_{\ff t,0} = 1 /(\omega + \mu - \ff t)$ for the free (non-interacting) Green's
function which exhibits the dependence on the model parameters but suppresses the 
frequency dependence.
Dyson's equation then reads as
$\ff G_{\ff t, \ff U} = 1 / (\ff G^{-1}_{\ff t,0} - \ff \Sigma_{\ff t, \ff U})$.

The Luttinger sum rule \cite{LW60,Lut60} states that 
\begin{equation}
  \langle N \rangle = 2 \sum_{\ff k} \Theta (G_{\ff k}(0))
\label{eq:lutt}  
\end{equation}
where $N = \sum_{i\sigma} n_{i\sigma}$ is the particle-number operator, $\langle N \rangle$
its ($T=0$) expectation value, and $\Theta$ the Heavyside step function.
The factor $2$ accounts for the two spin directions.
Since $G_{\ff k}(0)^{-1} = \mu - \varepsilon(\ff k) - \Sigma_{\ff k}(0)$, the sum
gives the number of $\ff k$ points enclosed by the interacting Fermi 
surface which, for $L\to \infty$, is defined via 
$\mu - \varepsilon(\ff k) - \Sigma_{\ff k}(0) = 0$.
In the thermodynamic limit the sum rule therefore equates the average particle number 
with the Fermi-surface volume (apart from a factor $(2\pi)^D/L$).
Note that, as $\Theta(G_{\ff k}(0))=\Theta(1/G_{\ff k}(0))$, the sum rule \refeq{lutt} 
also includes the so-called Luttinger volume \cite{Dzy03} 
which (for $L\to \infty$) is enclosed by the {\em zeros} of $G_{\ff k}(0)$.

The standard proof of the sum rule can be found in Ref.\ \onlinecite{LW60}.
It is based on diagrammatic perturbation theory to all orders which is used to construct 
the Luttinger-Ward functional $\Phi_{\ff U} [\ff G]$ as the sum of renormalized closed 
skeleton diagrams (see Fig.\ \ref{fig:phi}).
We emphasize that the original proof straightforwardly extends also to finite systems.
For $L < \infty$ the sum in \refeq{lutt} is discrete. 
Actually, the proof is performed for finite $L$ first, and the thermodynamic limit
(if desired) can be taken in the end. 
The limit $T \to 0$, on the other hand, is essential and is responsible for possible 
violations of the sum rule (see Sec.\ \ref{sec:vio}). 

Below we need an alternative but equivalent formulation of the sum rule. 
We start from the following (Lehmann) representation for the Green's function:
\begin{equation}
  G_{\ff k}(\omega) = \sum_m \frac{\alpha_m(\ff k)}{\omega + \mu - \omega_m(\ff k)} \: .
\end{equation}
Here, $\omega_m(\ff k)-\mu$ are the (real) poles and $\alpha_m(\ff k)$ the (real and positive)
weights.
For real frequencies $\omega$, it is then easy to verify the identity:
\begin{equation}
  \Theta(G_{\ff k}(\omega)) = \sum_m \Theta(\omega+\mu-\omega_m(\ff k)) - \sum_n \Theta(\omega+\mu-\zeta_n(\ff k))
\label{eq:th}
\end{equation}
where $\zeta_n(\ff k)-\mu$ is the $n$-th (real) zero of the Green's function, i.e.\
$G_{\ff k}(\zeta_n(\ff k)-\mu)=0$.

For temperature $T=0$ we have 
$\langle N \rangle = 2\sum_{\ff k} \int_{-\infty}^0 d\omega (-1/\pi) \mbox{Im} G_{\ff k}(\omega+i0^+)$ 
and thus $\langle N \rangle = 2\sum_{\ff k} \sum_m \alpha_m(\ff k) \Theta(\mu-\omega_m(\ff k))$.
Hence, the Luttinger sum rule reads:
\begin{eqnarray}
&& 
  2\sum_{\ff k} \sum_m \alpha_m(\ff k) \Theta(\mu-\omega_m(\ff k))
\nonumber \\   
&&
  = 2\sum_{\ff k} 
  \left(
  \sum_m \Theta(\mu-\omega_m(\ff k)) - \sum_n \Theta(\mu-\zeta_n(\ff k))
  \right) \: .
\nonumber \\   
\label{eq:lutt1}  
\end{eqnarray}
This form of the sum rule is convenient for the discussion of finite systems
with $L<\infty$.

\section{Self-energy-functional theory and Luttinger sum rule}
\label{sec:sft}

Within the self-energy-functional theory (SFT), \cite{Pot03a,Pot03b,PAD03} the grand 
potential $\Omega$ is considered as a functional of the self-energy:
\begin{equation}
  \Omega_{\ff t, \ff U}[\ff \Sigma] 
  = 
  \Tr \ln \frac{1}{\ff G_{\ff t,0}^{-1} - \ff \Sigma} + F_{\ff U}[\ff \Sigma] 
  \: .
\label{eq:sef}  
\end{equation}
Here, the trace $\mbox{Tr}$ of a quantity $\ff A$ is defined as
$\Tr \ff A \equiv T \sum_n 2 \sum_{\ff k} e^{i\omega_n0^+} A_{\ff k}(i\omega_n)$
where $i \omega_n = i (2n+1) \pi T$ are the fermionic Matsubara frequencies, and
the functional $F_{\ff U}[\ff \Sigma]$ is the Legendre transform of the Luttinger-Ward
functional $\Phi_{\ff U}[\ff G]$.
The self-energy functional (\ref{eq:sef}) is stationary at the physical self-energy, 
$\delta \Omega_{\ff t, \ff U}[\ff \Sigma_{\ff t,\ff U}] / \delta \ff \Sigma = 0$, and,
if evaluated at the physical self-energy, yields the physical value for the grand
potential: $\Omega_{\ff t, \ff U}[\ff \Sigma_{\ff t,\ff U}] = \Omega_{\ff t, \ff U}
\equiv -T \ln \tr \exp(-\beta (H(\ff t,\ff U) - \mu N))$ where $\beta=1/T$.

Comparing with the self-energy functional
\begin{equation}
  \Omega_{\ff t', \ff U}[\ff \Sigma] 
  = 
  \Tr \ln \frac{1}{\ff G_{\ff t',0}^{-1} - \ff \Sigma} + F_{\ff U}[\ff \Sigma]
\end{equation}
of a reference system with the same interaction but a modified one-particle part, 
i.e.\ with the Hamiltonian $H(\ff t',\ff U)$, the not explicitly known but only
$\ff U$-dependent functional $F_{\ff U}[\ff \Sigma]$ can be eliminated:
\begin{equation}
  \Omega_{\ff t, \ff U}[\ff \Sigma] 
  = 
  \Omega_{\ff t', \ff U}[\ff \Sigma] 
  +
  \Tr \ln \frac{1}{\ff G_{\ff t,0}^{-1} - \ff \Sigma}
  -
  \Tr \ln \frac{1}{\ff G_{\ff t',0}^{-1} - \ff \Sigma} \: .
\label{eq:sef1}
\end{equation}
An approximation is constructed by searching for a stationary point of the self-energy
functional on the subspace of trial self-energies spanned by varying the one-particle
parameters $\ff t'$:
\begin{equation}
  \frac{\partial \Omega_{\ff t, \ff U}[\ff \Sigma_{\ff t',\ff U}] }{\partial \ff t'} = 0 \: .
\end{equation}
Inserting a trial self-energy into Eq.\ (\ref{eq:sef1}) yields
\begin{equation}
  \Omega_{\ff t, \ff U}[\ff \Sigma_{\ff t',\ff U}] 
  =
  \Omega_{\ff t', \ff U}
  +
  \Tr \ln \frac{1}{\ff G_{\ff t,0}^{-1} - \ff \Sigma_{\ff t',\ff U}}
  -
  \Tr \ln \ff G_{\ff t',\ff U} \: .
\label{eq:om}  
\end{equation}
The decisive point is that the r.h.s.\ can be evaluated exactly for a reference 
system which is exactly solvable. 
Apart from the free Green's function $\ff G_{\ff t,0}$, it involves quantities of 
the reference system only.

This strategy to generate approximations has several advantages: 
(i) Contrary to the usual conserving approximations, the exact functional form of 
$\Omega_{\ff t, \ff U}[\ff \Sigma]$ is retained. 
Any approximation is therefore non-perturbative by construction.
On the level of one-particle excitations, macroscopic conservation laws are respected
as shown in Appendix \ref{sec:cons}.
(ii) With $\Omega_{\ff t, \ff U}[\ff \Sigma_{\ff t',\ff U}]$ evaluated at the stationary
point $\ff t' = \ff t'_{\rm s}$, an approximate but explicit expression for a 
thermodynamical potential is provided. 
As all physical quantities derive from this potential, the approximation is 
thermodynamically consistent in itself (see Ref.\ \onlinecite{AAPH06a} for details).
(iii) As different reference systems generate different approximations, the SFT provides 
a unifying framework that systematizes a class of ``dynamic'' approximations (see
Refs.\ \onlinecite{Pot05,PB07} for a discussion).

In the following we discuss the question whether or not a dynamic approximation 
respects the Luttinger sum rule.
For this purpose consider first the $\Tr \ln( \cdots)$ terms in Eq.\ (\ref{eq:om}).
These can be evaluated using the analytical and causal properties of the Green's 
functions as described in Ref.\ \onlinecite{Pot03b} (see Eq.\ (4) therein).
Using $- T \ln ( 1 + \exp(-\omega/T)) \to \omega \Theta(-\omega)$ for $T \to 0$ yields:
\begin{eqnarray}
&&
  \Tr \ln \frac{1}{\ff G_{\ff t,0}^{-1} - \ff \Sigma_{\ff t',\ff U}}
\nonumber \\
&& 
  = 2 \sum_{\ff k} \sum_m (\omega_m(\ff k) - \mu) \Theta(\mu - \omega_m(\ff k))
\nonumber \\
&& 
  - 2 \sum_{\ff k} \sum_n (\zeta_n(\ff k) - \mu) \Theta(\mu - \zeta_n(\ff k)) \: .
\label{eq:trlng}
\end{eqnarray}
Analogously, we have
\begin{eqnarray}
  \Tr \ln \ff G_{\ff t',\ff U}
  &=& 
  2 \sum_{\ff k} \sum_m (\omega'_m(\ff k) - \mu) \Theta(\mu - \omega'_m(\ff k))
\nonumber \\
&& 
  - 2 \sum_{\ff k} \sum_n (\zeta_n(\ff k) - \mu) \Theta(\mu - \zeta_n(\ff k)) \: .
\nonumber \\
\label{eq:trlngs}
\end{eqnarray}
Note that the reference system is always assumed to be in the same macroscopic state
as the original system, i.e.\ it is considered at the same temperature and, more 
importantly here, at the same chemical potential $\mu$.
Furthermore, it has been used that, by construction of the approximation, the self-energy 
and hence its poles at $\zeta_n(\ff k)-\mu$ are the same for both, the original and the 
reference system.
This implies that the second terms on the r.h.s.\ of Eq.\ (\ref{eq:trlng}) 
and (\ref{eq:trlngs}), respectively, cancel each other in Eq.\ (\ref{eq:om}).
Finally, a (large but) finite system ($L<\infty$) and a finite reference system are
considered.
Hence, the set of poles of the Green's function and of the self-energy as well as sums 
over $\ff k$ are discrete and finite.

Taking the $\mu$ derivative on both sides of Eq.\ (\ref{eq:om}) then yields:
\begin{eqnarray}
  \frac{\partial \Omega_{\ff t, \ff U}[\ff \Sigma_{\ff t', \ff U}]}{\partial \mu}
  &=&
  \frac{\partial \Omega_{\ff t', \ff U}}{\partial \mu}
  - 2 \sum_{\ff k} \sum_m \Theta(\mu - \omega_m(\ff k))
  \nonumber \\
  &+& 2 \sum_{\ff k} \sum_m \Theta(\mu - \omega'_m(\ff k)) \: .
\end{eqnarray}
Here we have assumed the ground state of the reference system to be non-degenerate with 
respect to the particle number.
From the (zero-temperature) Lehmann representation \cite{FW71} it is then obvious that,
within a subspace of fixed particle number, 
the $\mu$-dependence of the Green's function is the same as its $\omega$-dependence,
i.e.\ $\ff G(\omega) = \widetilde{\ff G}(\omega + \mu)$ with a $\mu$-independent function
$\widetilde{\ff G}$.
Via the Dyson equation of the reference system, this property can also be inferred for
the self-energy and, via the Dyson equation of the original system, for the (approximate)
Green's function of the original system. 
Consequently, the poles of $(\ff G_{\ff t,0}^{-1} - \ff \Sigma_{\ff t',\ff U})^{-1}$ and
of $\ff G_{\ff t',\ff U}$ are linearly dependent on $\mu$, i.e.\ $\omega_m(\ff k)$ and
$\omega'_m(\ff k)$ in Eqs.\ (\ref{eq:trlng}) and (\ref{eq:trlngs}) are independent of
$\mu$.

We once more exploit the fact that the self-energy of the original system is identified
with the self-energy of the reference system.
Using Eq.\ (\ref{eq:th}) one immediately arrives at
\begin{equation}
  \langle N \rangle = \langle N \rangle' + 2 \sum_{\ff k} \Theta(G_{\ff k}(0))
  - 2 \sum_{\ff k} \Theta(G'_{\ff k}(0)) \: .
\label{eq:res}
\end{equation}
This is the final result:
The Luttinger sum rule for the original system,
Eq.\ (\ref{eq:lutt}), is satisfied if and only if it is satisfied for the reference
system, i.e.\ if $\langle N \rangle'=2 \sum_{\ff k} \Theta(G'_{\ff k}(0))$.

A few remarks are in order.
For the reference system, the status of the Luttinger sum rule is that of a general 
theorem (as long as the general proof is valid); $\langle N \rangle'$ and 
$G'_{\ff k}(0)$ represent exact quantities.
The above derivation shows that the theorem is ``propagated'' to the original system 
{\em irrespective of the approximation that is constructed within the SFT}.
This propagation also works in the opposite direction. 
Namely, a possible violation of the exact sum rule for the reference system would 
imply a violation of the sum rule, expressed in terms of approximate quantities, 
for the original system. 

Eq.\ (\ref{eq:res}) holds for any choice of $\ff t'$. 
Note, however, that stationarity with respect to the variational parameters $\ff t'$
is essential for the thermodynamical consistency of the approximation.
In particular, consistency means that the average particle number
$\langle N \rangle = - \partial \Omega_{\ff t, \ff U}[\ff \Sigma_{\ff t',\ff U}]/\partial \mu$ 
on the l.h.s.\ can be obtained as the trace of the Green's function.
Stationarity is thus necessary to get the sum rule in the form (\ref{eq:lutt1}). 

There are no problems to take the thermodynamic limit (if desired) on both sides
of Eq.\ (\ref{eq:res}) (after division of both sides by the number of sites $L$).
The $\ff k$ sums turn into integrals over the unit cell of the reciprocal lattice.
For a $D$-dimensional lattice the $D-1$-dimensional manifolds of $\ff k$ points with 
$G_{\ff k}(0)=\infty$ or $G_{\ff k}(0)=0$ form Fermi or Luttinger surfaces, respectively.

For the above derivation, translational symmetry has been assumed for both, the original 
as well as the reference system.
Nothing, however, prevents us from repeating the derivation in case of systems with
reduced (or completely absent) translational symmetries.
One simply has to re-interprete the wave vector $\ff k$ as an index which, combined 
with $m$, refers to the elements of the diagonalized Green's function matrix $\ff G$.
The exact sum rule, Eq.\ (\ref{eq:lutt1}), generalizes accordingly.
The result (\ref{eq:res}) remains valid (with the correct interpretation of $\ff k$) 
for an original system with reduced translational symmetries.
It is also valid for the case of a translationally symmetric original Hamiltonian where, 
due to the choice of a reference system with reduced translational symmetries, the 
symmetries of the (approximate) Green's function of the original system are 
(artificially) reduced.
A typical example is the variational cluster approximation (VCA) where the reference system 
consists of isolated clusters of finite size.

\section{Two-site dynamical-impurity approximation}
\label{sec:two}

While the Hartree-Fock approximation may be considered as the most simple weak-coupling
$\Phi$-derivable approximation, the most simple non-perturbative $\Phi$-derivable  
approximation is given by the dynamical-impurity approximation (DIA).
This shall be demonstrated in the following for the single-band Hubbard model (\ref{eq:hub})
as the original system to be investigated. 
The DIA is generated by a reference system consisting of a decoupled set of single-impurity
Anderson models with a finite number of sites $n_{\rm s}$ and is known \cite{Pot03a} to 
recover the dynamical mean-field theory in the limit $n_{\rm s} \to \infty$.
As long as the Luttinger sum rule holds for the single-impurity reference system, the DIA 
must yield a one-particle Green's function and a self-energy respecting the sum rule.

The Hamiltonian of the reference system is $H(\ff t',\ff U) = \sum_{i=1}^L H'_i$ with
\begin{eqnarray}
   H'_i &=& \sum_\sigma \varepsilon_{0} c^\dagger_{i\sigma} c_{i\sigma} 
   + \frac{U}{2} \sum_\sigma n_{i\sigma} n_{i-\sigma}
   \nonumber \\
   &+& \sum_{k=2}^{n_{\rm s}} \sum_\sigma \varepsilon_{k} a^\dagger_{ik\sigma} a_{ik\sigma} 
   + \sum_{k=2}^{n_{\rm s}} \sum_\sigma V_{k} (a^\dagger_{ik\sigma} c_{i\sigma} + \mbox{h.c.})   
   \: .
   \nonumber \\
\end{eqnarray}
For a homogeneous phase, the variational parameters 
$\ff t' = (\{\varepsilon_0^{(i)},\varepsilon_{k}^{(i)},V_k^{(i)}\})$
can be assumed to be independent of the site index $i$:
$\varepsilon_0 \equiv \varepsilon_0^{(i)}$, $\varepsilon_{k} \equiv \varepsilon_{k}^{(i)}$,
$V_k \equiv V_k^{(i)}$.
For the sake of simplicity, we consider the two-site DIA ($n_{\rm s}=2$), i.e.\ a single 
bath site per correlated site only. 
In this case there are three independent variational parameters only: 
the on-site energies of the correlated and of the bath site, $\varepsilon_0$ and 
$\varepsilon_{\rm c} \equiv \varepsilon_{k=2}$, respectively, as well as the hybridization
strength $V\equiv V_{k=2}$.
As the reference system consists of replicated identical impurity models which are spatially 
decoupled, the trial self-energy is local and site-independent, 
$\Sigma_{ij}(\omega)=\delta_{ij}\Sigma(\omega)$. 

Calculations have been performed for the Hubbard model with a one-particle dispersion
$\varepsilon(\ff k) = L^{-1} \sum_{ij} e^{-i \ff k (\ff R_i - \ff R_j)} t_{ij}$
such that the density of one-particle energies $D(\varepsilon)$ is semi-elliptic. 
For $|\varepsilon| \le W/2$, 
\begin{equation}
  D(\varepsilon) = \frac{1}{L} \sum_{\ff k} \delta(\varepsilon-\varepsilon(\ff k))
  = \frac{8}{\pi W^2} \sqrt{(W/2)^2 - \varepsilon^2} \: . 
\end{equation}
The free band width is set to $W=4$.
This serves as the energy scale.

\begin{figure}[t]
  \includegraphics[width=0.7\columnwidth]{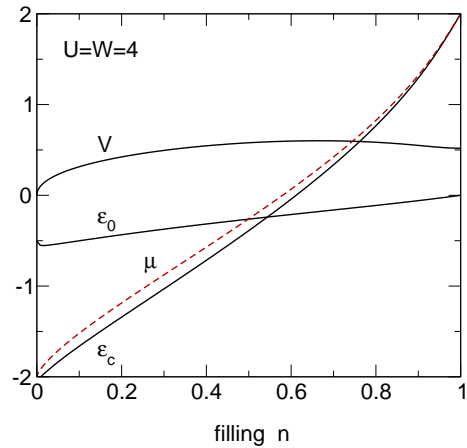}
\caption{
Filling dependence of the variational parameters at their respective optimized
values and of the chemical potential.
Calculations for the Hubbard model with a semi-elliptical free density of states 
of band width $W=4$ and interaction strength $U=W=4$ using the two-site DIA.
}
\label{fig:para}
\end{figure}

The computation of the SFT grand potential is performed as described in 
Ref.\ \onlinecite{Pot03b}. 
Stationary points of the resulting function $\Omega(\varepsilon_0,\varepsilon_{\rm c},V)
\equiv \Omega_{\ff t, \ff U}[\ff \Sigma_{\varepsilon_0,\varepsilon_{\rm c},V}]$ are
obtained via iterated linearizations of its gradient. 
There is a unique non-trivial stationary point (with $V \ne 0$). 
Fig.\ \ref{fig:para} shows the variational parameters at this point as functions of 
the filling $n$. 
For the entire range of fillings, the ground state of the reference system lies in the
invariant subspace with $N_{\rm tot} = \sum_\sigma ( c_{i\sigma}^\dagger c_{i\sigma} +
a_{i\sigma}^\dagger a_{i\sigma}) = 2$.
The parameters as well as the chemical potential are smooth functions of $n$.
We have checked that the thermodynamical consistency condition
$n = - L^{-1} \partial \Omega / \partial \mu = \int_{-\infty}^0 \rho(\omega) d\omega$
is satisfied within numerical accuracy.
Here
\begin{equation}
  \rho(\omega) = D(\omega + \mu - \Sigma(\omega))
\label{eq:dos}
\end{equation}
is the interacting local density of states (DOS).

At half-filling the values of the optimized on-site energies are consistent with 
particle-hole symmetry. 
With $\varepsilon_0 - \mu = -U/2$ and $\varepsilon_{\rm c} - \mu = 0$ the reference
system is in the Kondo regime with a well-formed local moment at the correlated site.
The finite hybridization strength $V$ leads, for $U=W$, to a finite DOS
$\rho(\omega=0)>0$ and thus to a metallic Fermi liquid as it is expected for the 
Hubbard model within a (dynamical) mean-field description.
Due to the simple structure of the self-energy generated by the two-site reference
system, however, quasi-particle damping effects are missing.

\begin{figure}[t]
  \includegraphics[width=0.8\columnwidth]{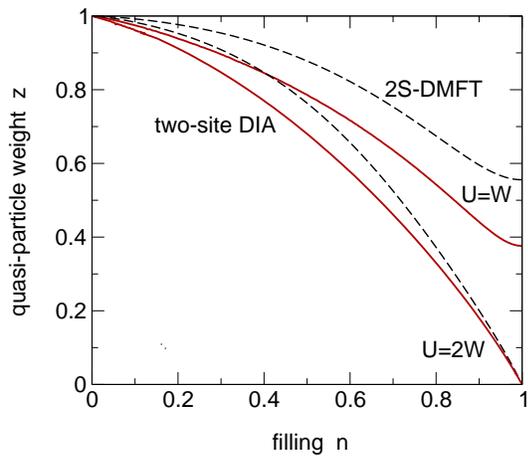}
\caption{
Quasi-particle weight $z$ as a function of the filling within the two-site DIA
(full lines) and the two-site DMFT \cite{Pot01} (dashed lines).
Calculations for $U=W$ and $U=2W$.
}
\label{fig:zofn}
\end{figure}

Decreasing the filling from $n=1$ to $n=0$ drives the reference system more and more
out of the Kondo regime.
While $\varepsilon_{\rm c}$ stays close to the chemical potential, the on-site energy
of the correlated site $\varepsilon_{0}$ crosses $\mu$ close to quarter filling 
and lies above $\mu$ eventually.
Note that $\varepsilon_{0}=0$ within the DMFT, i.e.\ for $n_{\rm s}\to \infty$, while
for finite $n_{\rm s}$ there is a clear deviation from $\varepsilon_{0}=0$ which is 
necessary to ensure thermodynamical consistency.
For fillings very close to $n=0$, the grand potential
$\Omega_{\ff t, \ff U}[\ff \Sigma_{\varepsilon_0,\varepsilon_{\rm c},V}]$
becomes almost independent of $\ff \Sigma$.
This implies that it becomes increasingly difficult to locate the stationary point
with the numerical algorithm used. 
The slight upturn of $\varepsilon_0$ below $n=0.01$ (see Fig.\ \ref{fig:para}) might 
be a numerical artifact.

It is instructive to compare the parameters with those of the two-site DMFT (2S-DMFT). 
\cite{Pot01}
The 2S-DMFT is a simplified version of the DMFT where a mapping onto the two-site 
single impurity Anderson model is achieved by means of a simplified self-consistency 
equation.
Assuming $\varepsilon_{0}=0$ as in the full DMFT, there are two parameters left 
($\varepsilon_{\rm c}$ and $V$) which are fixed by considering the first non-trivial
order in the low- and in the high-frequency expansion of the self-energy and the 
Green's function in the DMFT self-consistency equation.
Although being well motivated, this approximation is essentially {\em ad hoc}. 
One therefore has to expect that the 2S-DMFT is thermodynamically inconsistent and 
exhibits a violation of Luttinger's sum rule.
A comparison of the DIA for $n_{\rm s}=2$ with the 2S-DMFT is thus ideally suited to 
demonstrate the advantages gained by constructing approximations within the variational
framework of the SFT.

First of all, there are differences in fact.
At half-filling the 2S-DMFT predicts the hybridization to be somewhat larger than the
two-site DIA while the value for $\varepsilon_{\rm c}$ is again fixed by particle-hole
symmetry. 
Deviations grow with decreasing filling. 
Contrary to the two-site DIA, $V$ monotonously increases and is larger in the entire
filling range, $\varepsilon_{0}=0$ by construction, and $\varepsilon_{\rm c}$ even 
diverges for $n\to 0$ within the 2S-DMFT (see Ref.\ \onlinecite{Pot01}).
On the other hand, the system is essentially uncorrelated in the limit $n\to 0$.
Strong differences in the parameters, which enter the self-energy only, therefore do 
not necessarily imply strongly different physical quantities.
This is demonstrated by Fig.\ \ref{fig:zofn} which shows the quasi-particle weight 
calculated via 
\begin{equation}
  z = \left( 1-\frac{d \Sigma (\omega=0)}{d\omega} \right)^{-1} \: 
\end{equation}
as a function of the filling. 
While there are obvious differences when comparing the results from the two-site DIA
with those of the 2S-DMFT, the qualitative trend of $z$ is very similar in both 
approximations. 
Both approximations also compare well with the full DMFT: 
There is a quadratic behavior of $z(n)$ for $n\to 1$ in the Fermi-liquid phase ($U=W$)
and a linear trend when approaching the Mott phase ($U=2W$).
The critical interaction strength for the Mott transition is found to be 
$U_{\rm c} \approx 1.46 W$ for the two-site DIA and $U_c=1.5 W$ within the 2S-DMFT.
For details on the Mott transition see Refs.\ \onlinecite{Pot03b,Pot01}.

In case of a local and site-independent self-energy, the Luttinger sum rule can be 
written in the form \cite{MH89c}
\begin{equation}
   \mu = \mu_0 + \Sigma(\omega=0) \: ,
\label{eq:luttmu}
\end{equation}
where $\mu_0$ is the chemical potential of the free ($U=0$) system at the same particle 
density.
Eq.\ (\ref{eq:luttmu}) implies that not only the enclosed volume but also the shape of 
the Fermi surface remains unchanged when switching on the interaction. 
Using Eq.\ (\ref{eq:dos}) this immediately implies \cite{MH89c}
\begin{equation}
  \rho(0) = D(\mu_0) = \rho_0(0) \: ,
\label{eq:lutt2}
\end{equation}
i.e., in case of a correlated metal, the value of the interacting local density of states 
at $\omega=0$ is independent of $U$ and thus fixed to the value of the density of states 
of the non-interacting system at the same filling.

\begin{figure}[t]
  \includegraphics[width=0.85\columnwidth]{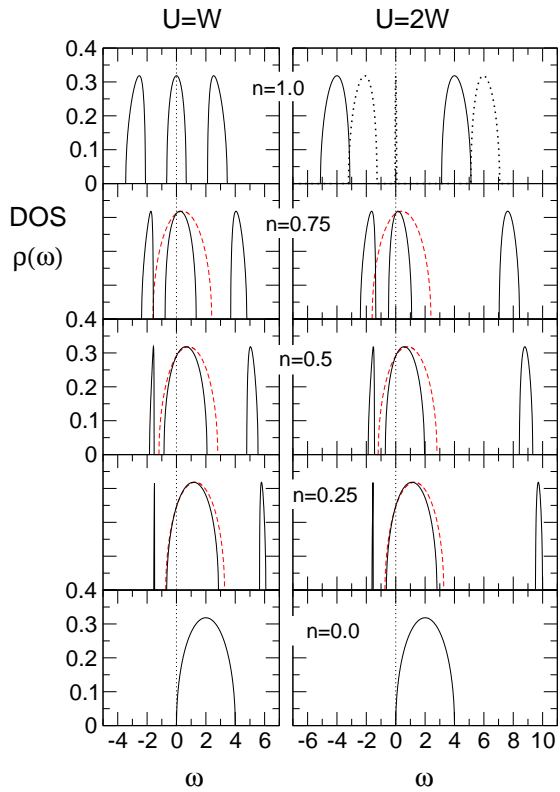}
\caption{
Interacting local density of states $\rho(\omega)$ (solid lines) for different fillings 
as indicated.
Calculations using the two-site DIA for $U=W$ (left) and $U=2W$ (right).
For $n=0.25$, $n=0.5$ and $n=0.75$ the non-interacting DOS $\rho_0(\omega)$ is shown 
for comparison (dashed lines). 
Note that $\rho(0)=\rho_0(0)$.
The dotted line for $U=2W$ in the top panel is the DOS for $n=0.99$.
}
\label{fig:dos}
\end{figure}

\begin{figure}[t]
  \includegraphics[width=0.8\columnwidth]{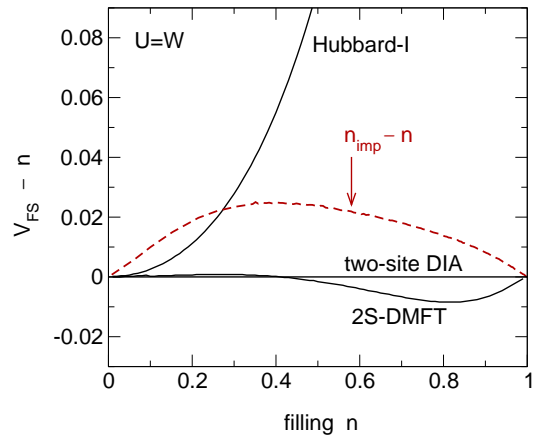}
\caption{
Numerical results for the difference between the volume enclosed by the Fermi surface 
$V_{\rm FS}$ and the filling $n$ as a function of $n$ for $U=W=4$. 
The Luttinger sum rule ($V_{\rm FS} - n = 0$) is exactly respected by the two-site
DIA. Results for the 2S-DMFT and the Hubbard-I approximation are shown for comparison. 
Dashed line: difference between the filling $n$ and the average occupation of the 
correlated (impurity) site in the reference system at stationarity for the two-site 
DIA.
}
\label{fig:lsr}
\end{figure}

\begin{figure}[b]
  \includegraphics[width=0.75\columnwidth]{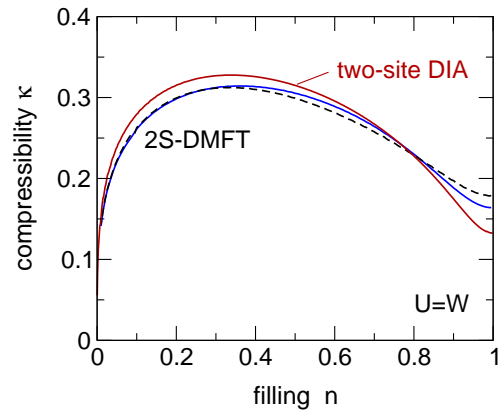}
\caption{
Filling dependence of the compressibility $\kappa$ for $U=W$ as obtained within
the 2S-DMFT via $\kappa = \partial n /\partial \mu$ (solid line) and via a general
Fermi-liquid relation (Eq.\ (\ref{eq:kap}), dashed line).
Using the two-site DIA identical results are obtained for both cases.
}
\label{fig:kappa}
\end{figure}

The interacting and the non-interacting DOS are plotted in Fig.\ \ref{fig:dos}
for different fillings and for $U=W$ and $U=2W$. 
The impurity self-energy of the two-site reference system is an analytical function of 
$\omega$ except for two first-order poles on the real axis. 
Via Eq.\ (\ref{eq:dos}) this two-pole structure implies that the DOS consists of three
peaks the form of which is essentially given by the non-interacting DOS.
At half-filling the three peaks are easily identified as the lower and the upper 
Hubbard band and the quasi-particle resonance as it is characteristic for a (dynamical) 
mean-field description. \cite{GKKR96}
For $U=W$ the resonance still has a significant weight.
The weight decreases upon approaching the critical interaction, and the resonance has
disappeared in the Mott insulator for $U=2W$.
Hole doping of the Mott insulator is accomplished by the reappearance of the resonance 
at $\omega=0$ which preempts the creation of holes in the lower Hubbard band. \cite{FKM95}
As can be seen in the spectrum for $n=0.99$ in the top panel (dotted line), the
quasi-particle resonance appears within the Mott-Hubbard gap.
With decreasing filling, the upper Hubbard band gradually shifts to higher excitation
energies and loses weight. 
This weight is transfered to the low-energy part of the spectrum.
For lower fillings where the Kondo regime has been left, one would actually expect that 
the quasi-particle resonance disappears by merging with the lower Hubbard band. 
This, however, cannot be described with the simple two-pole structure of the self-energy.
One therefore should interprete the gap around $\omega = -1$ at $n=0.25$ as an artifact 
of the approximation. 
Furthermore, the widths of the Hubbard bands is considerably underestimated as damping
effects are missing completely.
The filling-dependent spectral-weight transfer across the Hubbard gap as well as the
energy positions of the main peaks, however, are in overall agreement with general 
expectations. \cite{HL67,Hub63}

It is worth emphasizing that this simple two-site dynamical-impurity approximation 
exactly fulfills the Luttinger sum rule.
In Fig.\ \ref{fig:dos} this can be seen by comparing with the DOS of the non-interacting
system (dashed lines).
The non-interacting DOS cuts the interacting one at $\omega=0$ which shows that 
Eq.\ (\ref{eq:lutt2}) is satisfied.
Note that this is trivial for $n=1$ as this is already enforced by particle-hole 
symmetry.
Off half-filling, however, the pinning of the DOS to its non-interacting value at 
$\omega=0$ is a consequence of $\Phi$-derivability and thereby a highly non-trivial 
feature.

In contrast, the 2S-DMFT does show a violation of Luttinger's sum rule which, however, 
must be attributed to the {\em ad hoc} nature of the approximation. 
Fig.\ \ref{fig:lsr} shows the difference between the volume enclosed by the Fermi surface
\begin{equation}
V_{\rm FS} = \frac{2}{L} \sum_{\ff k} \Theta(\mu -\varepsilon(\ff k) - \Sigma(0)) 
= 2 \int_{-\infty}^0 ß\!\!d\varepsilon D (\varepsilon + \mu  - \Sigma(0))
\end{equation}
and the filling $n$ as a function of the filling. 
As can be seen, there is an artificial violation of the sum rule for the 2S-DMFT which
is of the order of a few per cent while for the $\Phi$-derivable two-site DIA the sum
rule is fully respected.
Note that, unlike the DMFT and also unlike the simplified 2S-DMFT, the two-site DIA 
predicts a filling which slightly differs from the average occupation of the correlated
impurity site in the reference system (see dashed line in Fig.\ \ref{fig:lsr}).
For a finite number of bath sites $n_{\rm s}$ this appears to be necessary to fulfill 
the Luttinger sum rule.
The figure also shows the result obtained within the Hubbard-I approximation. \cite{Hub63} 
Here a very strong (artificial) violation of up to 100 \% (for $n$ close to half-filling) 
is obtained.
This should be considered as a strong drawback which is typical for uncontrolled
mean-field approximations. 

There are more relations which, analogously to the Luttinger sum rule, can be derived
by means of perturbation theory to all orders \cite{Lut60} in the exact theory and which
are respected by weak-coupling conserving approximations.
For example, the compressibility, defined as $\kappa = \partial n /\partial \mu$, can be 
shown to be related to the interacting DOS and the self-energy at the Fermi edge via
\begin{equation}
  \kappa = 2 \rho(0) \left(1 - \frac{\partial \Sigma(0)}{\partial \mu} \right) \: .
\label{eq:kap}  
\end{equation}
Fig.\ \ref{fig:kappa} shows that for the 2S-DMFT it makes a difference whether $\kappa$ is 
calculated as the $\mu$-derivative of the filling or via Eq.\ (\ref{eq:kap}).
Again, this must be attributed to the fact that the 2S-DMFT is not a $\Phi$-derivable
approximation. 
Contrary, the two-site DIA does respect the general Fermi-liquid property (\ref{eq:kap})
and thus yields the same result in both cases (see Fig.\ \ref{fig:kappa}).

\section{Violation of Luttinger's sum rule in finite systems}
\label{sec:vio}

The preceding section has demonstrated that the two-site DIA satisfies the Luttinger
sum rule. 
According to Eq.\ (\ref{eq:res}), we can conclude that the Luttinger sum rule
must hold for the corresponding reference system, i.e.\ for the two-site single-impurity 
Anderson model.
Of course, this can be verified more directly by evaluating Eq.\ (\ref{eq:lutt1}).
In case of a finite system or a system with reduced translational symmetries, the Green's
function is a matrix with elements $G_{\alpha\beta}(\omega)$ where $\alpha$ refers to 
the one-particle basis states, and the Luttinger sum rule reads:
\begin{equation}
  \sum_{k,m} \alpha^{(k)}_m \Theta(\mu-\omega_m^{(k)})
  = 
  \sum_{k,m} \Theta(\mu-\omega^{(k)}_m) - \sum_{k,n} \Theta(\mu-\zeta^{(k)}_n)
  \: .
\label{eq:lutt3}  
\end{equation}
Here the index $k$ labels the elements of the diagonalized Green's function, i.e.\
Eq.\ (\ref{eq:lutt1}) is generalized by replacing $(\ff k,\sigma) \to k$.
In case of an impurity model, Eq.\ (\ref{eq:lutt3}) actually represents the Friedel
sum rule. \cite{LA61,Lan66}
For the two-site single-impurity Anderson model, the different one-particle excitation 
energies $\omega_m^{(k)}-\mu$, the zeros of the Green's function $\zeta_n^{(k)}-\mu$ 
and the weights $\alpha^{(k)}_m$ are easily determined by full diagonalization.
We find that Eq.\ (\ref{eq:lutt3}) is satisfied in the entire parameter space
(except for $V=0$, see below).

Note that a violation of the sum rule occurs when, as a function of a model parameter $x$, 
a zero of the Green's function crosses $\omega=0$ for $x=x_c$.
At $x_c$ the number of negative zeros counted by the second term on the r.h.s.\ changes 
by one while the first term as well as the l.h.s.\ remain constant since (unlike a pole) 
a zero of the Green's function is generically not connected with a change of the ground 
state (level crossing).
This implies that the sum rule would be violated for $x<x_c$ or for $x>x_c$.

The case $V=0$ is exceptional.
Within the two-site DIA this corresponds to the Mott insulator (see Fig.\ \ref{fig:dos}, 
topmost panel for $U=2W$).
For $V=0$ the reference system consists of two decoupled sites, and the Green's function 
becomes diagonal in the site index. 
There is no zero of the local Green's function corresponding to the uncorrelated site. 
We can thus concentrate on the correlated site where the local Green's function 
exhibits a zero at $\eta - \mu = \varepsilon_0 + U/2$.
In the sector with one electron at the correlated site 
($\varepsilon_0 < \mu < \varepsilon_0 +U$), the second term on the r.h.s.\ changes
by {\em two} at $\mu = \mu_c = \varepsilon_0 + U/2$ because of the two-fold degenerate
ground state. 
In this case Luttinger's sum rule in the form (\ref{eq:lutt3}) is violated for 
$\mu < \mu_c$ {\em and} for $\mu > \mu_c$.
This ``violation'', however, is a trivial one which immediately disappears if the
ground-state degeneracy is lifted by applying a weak field term, for example.

\begin{figure}[t]
  \includegraphics[width=0.8\columnwidth,angle=0]{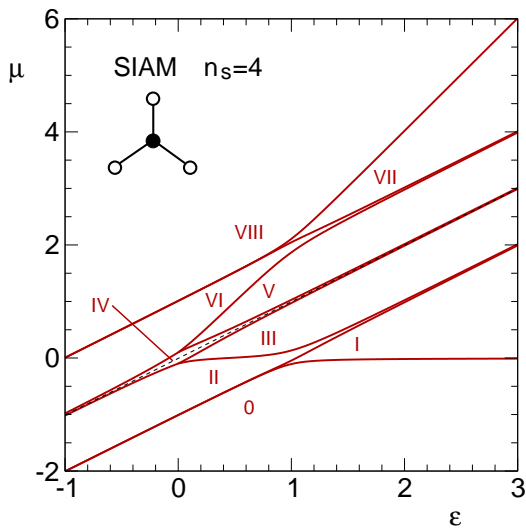}
\caption{
Phase diagram $\mu$ vs.\ $\varepsilon$ of the single-impurity Anderson model 
with $n_{\rm s}=4$ sites.
Total particle numbers are indicated by Roman figures.
Results have been obtained by full diagonalization for the following model 
parameters.
One-particle energies: $\varepsilon_0=0$ (correlated site), 
$\varepsilon_k = \varepsilon + (k-3)$ with $k=2,3,4$ (uncorrelated bath sites).
Hubbard interaction: $U=2\varepsilon$.
Hybridization strength: $V_k=0.1$ for $k=2,3,4$.
To lift Kramers degeneracy in case of an odd particle number, a weak (ferromagnetic) 
field of strength $b=0.001$ is coupled to the local spins.
The dashed line marks the particle-hole symmetric case.
Luttinger's sum rule is found to be satisfied in the entire parameter space.
}
\label{fig:m20}
\end{figure}

\begin{figure}[t]
  \includegraphics[width=0.8\columnwidth]{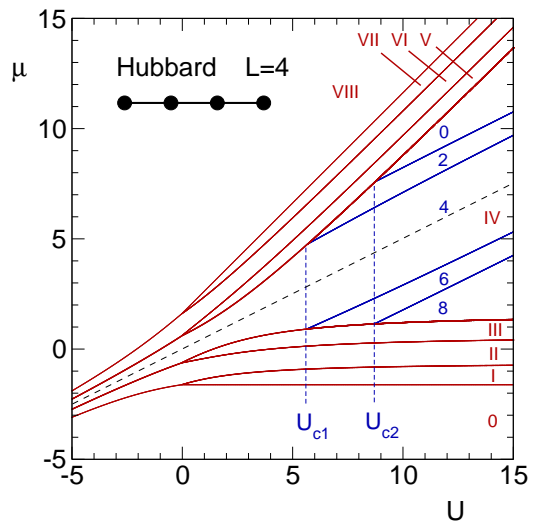}
\caption{
Phase diagram $\mu$ vs.\ $U$ of the Hubbard model with $L=4$ sites (open chain) 
as obtained by full diagonalization.
Nearest-neighbor hopping $t=-1$.
A weak (ferromagnetic) field of strength $b=0.01$ is applied to lift Kramers
degeneracy.
The dashed line marks the particle-hole symmetric case.
Particle numbers (l.h.s.\ of Eq.\ (\ref{eq:lutt3})) are indicated by Roman figures.
R.h.s.\ of Eq.\ (\ref{eq:lutt3}): Arabic figures.
Luttinger's sum rule is found to be violated for sufficiently strong $U$.
$U_{c1}$, $U_{c,2}$: critical interactions.
}
\label{fig:m22}
\end{figure}

Fig.\ \ref{fig:m20} shows a phase diagram of the single-impurity Anderson model
with $n_{\rm s}=4$ sites as obtained by full diagonalization.
The diagram covers the entire range of the total particle
number $N = \sum_\sigma \langle c_\sigma^\dagger c_{\sigma} \rangle
+ \sum_\sigma \sum_{k=2}^{n_{\rm s}} \langle a_{k\sigma}^\dagger a_{k\sigma} \rangle$ 
from $N = 0$ to $N = 2 n_{\rm s} = 8$.
A non-degenerate ground state is enforced by applying a small but finite magnetic
field.
No violation of the Luttinger sum rule is found.
We have repeated the same calculation also for $n_{\rm s}=10$ using the 
Lanczos technique. \cite{LG93}
Again, the sum rule is found to be always satisfied (We have performed calculations
for different $U$ and bath parameters).
This might have been expected as the ($n_{\rm s}\to \infty$) Anderson model can generally 
be classified as a (local) Fermi liquid. \cite{Hew93}

The situation is less clear in the case of correlated lattice models such as 
the Hubbard or the $t$-$J$ model.
For two dimensions there are several numerical studies using
high-temperature expansion, \cite{PLS98}
quantum Monte-Carlo, \cite{GEH00}
extended DMFT, \cite{HRKW02,HRKW03}
and dynamical cluster approximation (DCA) \cite{MPJ02}
which indicate a violation in the strongly correlated metallic phase
close to half-filling.
For studies of large clusters or studies directly working in the thermodynamic limit, 
a definite conclusion on the validity of the sum rule is difficult to obtain as 
finite-temperature or artificial broadening effects etc.\ must be controlled numerically.
Contrary, full diagonalization of Hubbard clusters consisting of a few sites only can
provide exact results. 
While their direct relevance for the thermodynamic limit is less clear, it is important 
to note that reference systems with a finite number of sites or a finite number of 
correlated sites provide the basis for a number of cluster approaches within the 
SFT framework. 
Via Eq.\ (\ref{eq:res}) their properties are transferred to the approximate treatment
of lattice models in the thermodynamic limit.

The validity of Eq.\ (\ref{eq:lutt3}) has been checked for Hubbard clusters of 
different size and in different geometries.
The $\mu$ vs.\ $U$ phase diagram for an $L=4$-site open Hubbard chain with 
nearest-neighbor hopping in Fig.\ \ref{fig:m22} shows a representative example. 
Again, a small but finite field term is added to avoid a ground-state degeneracy.
As the chemical potential, for fixed $U$, is moved off the particle-hole symmetric 
point $\mu=U/2$ and exceeds certain critical values (red lines), the particle 
number $N$ [as obtained from the l.h.s.\ of Eq.\ (\ref{eq:lutt3})] changes from $N=L$ 
down to (up to) $N=0$ ($N=2L$).
A critical $\mu$ value indicates a change of the ground state (level crossing) that
is accompanied by a change of the ground-state particle number.
In the one-particle Green's function this is characterized by a pole $\omega_m^{(k)} - \mu$
crossing $\omega=0$.
The blue lines indicate those chemical potentials at which a zero of the 
Green's function $\zeta^{(k)}_n - \mu$ crosses $\omega=0$.
Whenever this happens the r.h.s.\ of Eq.\ (\ref{eq:lutt3}) changes while the l.h.s.\
is constant.
Fig.\ \ref{fig:m22} shows that this occurs several times in the $N=L$ sector. 
At the particle-hole symmetric point $\mu=U/2$ the Luttinger sum rule is obeyed while 
it is violated in a wide region of the parameter space corresponding to half-filling 
$N=L$.
However, a critical interaction strength $U_c$ turns out to be necessary. 
The value for $U_c$ strongly varies for different cluster sizes and geometries 
but has always been found to be positive and finite.
Note that for $L=4$ the sum rule is fulfilled for any particle number $N \ne L$.
Qualitatively similar results can be found for the $L=2$-site Hubbard cluster
where calculations can be done even analytically. 
Again, a violation of the sum rule is found in the half-filled sector beyond 
a certain critical $U$.

\begin{figure}[t]
  \includegraphics[width=0.7\columnwidth]{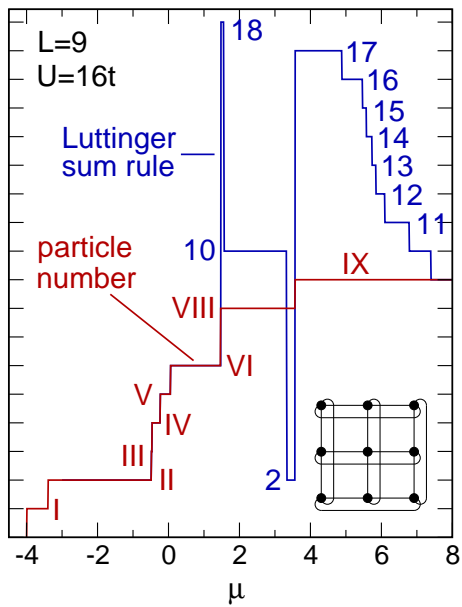}
\caption{
Ground-state particle number (red Roman figures, l.h.s.\ of Eq.\ (\ref{eq:lutt3})) and prediction
by the Luttinger sum rule (blue Arabic figures, r.h.s.\ of Eq.\ (\ref{eq:lutt3})) as functions of 
the chemical potential for a $L=9$-site Hubbard cluster with periodic boundary conditions. 
Arabic numbers are only given when different from Roman ones.
Calculations using the Lanczos method and a finite but small magnetic field and finite but
small on-site potentials to lift ground-state degeneracies.
}
\label{fig:hub9}
\end{figure}

\begin{figure}[t]
  \includegraphics[width=0.7\columnwidth]{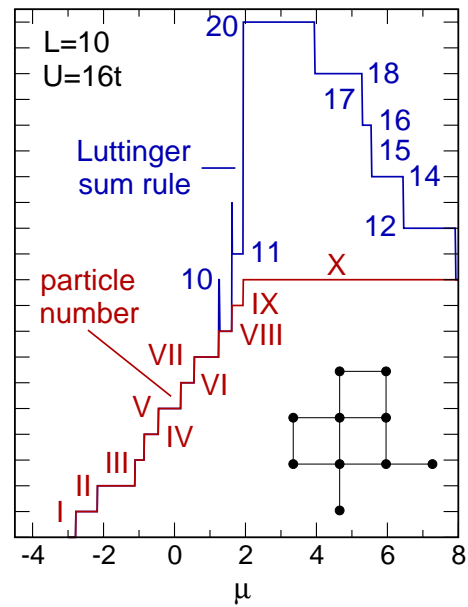}
\caption{
The same as Fig.\ \ref{fig:hub9} but for 10 sites. 
}
\label{fig:hub10}
\end{figure}

This has already been noticed by Rosch \cite{Ros06} and was used in combination 
with a strong-coupling expansion to argue that a violation of the sum rule 
generically occurs for a Mott insulator.
Stanescu et al.\ \cite{SPC07} have shown quite generally that the sum rule is 
fulfilled when particle-hole symmetry is present (the Luttinger surface is the
same as the Fermi surface of the non-interacting system) but violated in the 
Mott insulator away from particle-hole symmetry.
It is interesting to note that these arguments cannot be used to construct a
violation of the sum rule within DMFT or for a single-impurity Anderson model:
For an (almost) particle-hole symmetric case and model parameters describing 
a Mott insulator (within DMFT), an odd number of sites $n_{\rm s}$ (with 
$n_{\rm s}\to \infty$) must be considered and thus a magnetic field is needed 
to lift Kramers degeneracy.
Even an infinitesimal field, however, leads (at zero temperature) to a finite 
and even large polarization corresponding a well-formed but unscreened local 
moment.
This polarization is incomplete for any finite $U$ as the DMFT predicts a small
but finite double occupancy for a Mott insulator. 
Still there is a proximity to the fully polarized band insulator which finally 
results in a weakly correlated state and thus in a situation which is unlikely 
to show a violation of the sum rule. 

We have also considered Hubbard clusters with $L=9$ and $L=10$ sites by using the 
Lanczos technique. \cite{LG93}
Calculations have been performed for different Lanczos depths $l_{\rm max}$ to ensure 
that the results are independent of $l_{\rm max}$.
Fig.\ \ref{fig:hub9} displays an example for $L=9$ and a highly symmetric cluster 
geometry with periodic boundary conditions and a well-defined reciprocal space. 
To lift ground-state degeneracies resulting from spatial symmetries as well as the
Kramers degeneracy, small but finite on-site potentials and a small magnetic-field 
term are included in the cluster Hamiltonian. Fig.\ \ref{fig:hub10} shows an example 
for $L=10$ sites without any spatial symmetries. Kramers degeneracy for odd $N$ is 
removed by applying a small magnetic field.
With the figures we compare the expressions on the left-hand and the right-hand side 
of Eq.\ (\ref{eq:lutt3}). 
Obviously, the sum rule is respected in most cases. 
Violations are seen for half-filling $N=L$, i.e.\ in the ``Mott-insulating phase'', which
is consistent with Ref.\ \onlinecite{Ros06}. 
However, the sum rule is also violated in the ``metallic phase'' close to half-filling, 
namely for $N=L-1$ (Fig.\ \ref{fig:hub9}, $L=9$) and $N=L-1,L-2$ 
(Fig.\ \ref{fig:hub10}, $L=10$).
This nicely corresponds to the generally observed trend \cite{PLS98,GEH00,HRKW02,HRKW03,MPJ02}
for violations in the slightly doped metallic regime.
We have also verified that the sum rule is restored by lowering $U$.

Fig.\ \ref{fig:hub9} and \ref{fig:hub10} demonstrate that the sum rule is violated
in the {\em whole} $\mu$ range corresponding to $N=L-1$.
This is an important point as it shows that it is irrelevant whether the $T=0$ limit is
approached by holding $\langle N \rangle$ fixed and adjusting $\mu=\mu(T)$ or by fixing $\mu$ 
and let $\langle N \rangle=\langle N \rangle(T)$ be $T$-dependent.
A violation of the sum rule is found in both cases.

Kokalj and Prelov\u{s}ek \cite{KP07} have demonstrated that violations of the sum
rule can also be found for the $t$-$J$ model on a finite number of sites.
Our result provides an explicit example showing that not only for $t$-$J$ \cite{KP07} 
but also for Hubbard clusters a violation can be found when the chemical potential is 
set to $\mu=\lim_{T\to 0} \mu(T)$ with $\mu(T)$ obtained for given 
$\langle N \rangle = \mbox{const}$.
Anyway, the original proof \cite{LW60} does not depend on this choice for $\mu$ but 
appears to work for any $\mu$.

The results raise the question which assumptions used in the original proof of the theorem 
are violated or where the proof breaks down.
Note that the recently proposed alternative topological proof \cite{Osh00} assumes a 
Fermi-liquid state from the very beginning and thus cannot be applied to a 
finite system.
Using weak symmetry-breaking fields, a more or less trivial breakdown due to 
ground-state degeneracy has been excluded. 
An analysis of the ground state of the $L=2$ and $L=4$ Hubbard clusters which are accessible
with exact (analytical or numerical) methods has shown that, for model parameters where the
sum rule is violated, the interacting ground state can nevertheless be adiabatically 
connected to the non-interacting one. 
This excludes level crossing as a potential cause for the breakdown. 
While we cannot make a definite statement, 
it appears at least plausible that the violation of the sum rule results from a non-commutativity 
of two limiting processes, the infinite skeleton-diagram expansion and the limit $T\to 0$.

Using a functional-integral formalism, the Luttinger-Ward functional at finite $T$ can also 
be constructed in a {\em non-perturbative} way, i.e.\ avoiding an infinite summation of 
diagrams, as has been shown recently. \cite{Pot06b}
Formally, the Luttinger sum rule can be obtained by exploiting a gauge invariance of the 
Luttinger-Ward functional [see Ref.\ \onlinecite{Pot06b}]:
\begin{equation}
  \frac{\partial}{\partial (i\omega_n)} \Phi_{\ff U}[\ff G(i\omega_n)] = 0 \: .
\end{equation}
If at all, this invariance can only be shown for $T=0$ where $i\omega_n$ becomes a continuous
variable.
Unfortunately, the non-perturbative construction of $\Phi_{\ff U}$ requires a $T>0$ formalism. 
Hence, the validity of the sum rule depends on question whether the limit $T\to 0$ commutes
with the frequency differentiation. 
Necessary and sufficient conditions for this assumption are not easily worked out. 
An understanding of the main reason for the possible breakdown of the sum rule in finite
systems, very similar to the case of Mott insulators, is therefore not yet available 
(see also the discussion in Ref.\ \onlinecite{Ros06}).

\section{Conclusions}
\label{sec:con}

$\Phi$-derivable approximations are conserving, thermodynamically consistent and, for
$T=0$, formally respect certain non-trivial theorems such as the Luttinger sum rule. 
As the construction of the Luttinger-Ward functional $\Phi$ is by no means trivial and 
may conflict with the limit $T\to 0$ or different other limiting processes, however, 
the validity of the sum rule may be questioned.
Violations of the sum rule can be found in fact for the case of strongly correlated 
electron systems.
For Mott insulators and finite systems in particular, a breakdown is documented easily.

This implies that a general approximation for the spectrum of one-particle excitations
(of the one-particle Green's function) may violate the sum rule for {\em two} possible
reasons, namely because (i) the sum rule is violated in the exact theory, or (ii) the 
approximation generates an artificial violation.

Within the usual weak-coupling conserving approximations, such as the fluctuation-exchange 
approximation, the sum rule always holds as the formal steps in the general proof of the 
sum rule can be carried over to the approximation -- but with the important simplification 
of a limited class of diagrams.
This also implies that weak-coupling conserving approximations, when applied beyond the 
weak-coupling regime, might erroneously predict the sum rule to hold.

The present paper has focussed on {\em non-perturbative} conserving approximations.
Non-perturbative approximations, constructed within the framework of the self-energy-functional 
theory and referring to a certain reference system, are $\Phi$-derivable and consequently 
respect certain macroscopic conservation laws and are thermodynamically consistent. 
Whether or not the sum rule holds within the approximate approach, however, cannot be 
answered generally. 
We found that Luttinger's sum rule holds within an (SFT) approximation if and only if 
it holds exactly in the corresponding reference system. 

The reference system that leads to the most simple but non-trivial example for a 
non-perturbative conserving approximation consists of a single correlated and a single
bath site.
For this two-site system, we have found the sum rule to be valid in the entire parameter space.
Consequently, the resulting two-site dynamical-impurity approximation (DIA) -- opposed to 
more {\em ad hoc} approaches like the two-site DMFT -- fully respects the 
sum rule as could be demonstrated in different ways.
In view of the simplicity of the approximation this is a remarkable result.
Since the sum rule dictates the low-frequency behavior of the one-particle Green's function,
important mean-field concepts, such as the emergence of a quasi-particle resonance at the
Fermi edge, are qualitatively captured correctly, even away from the particle-hole symmetric
case.
This qualifies the two-site DIA for a quick but rough estimate of mean-field physics, 
including phases with spontaneously broken symmetries.

Full diagonalization and the Lanczos method have been employed to show that also the single-impurity
Anderson model with a finite number of $n_{\rm s}>2$ sites respects the sum rule. 
Consequently, this property is transferred to an $n_{\rm s}$-site DIA. 
For $n_{\rm s} \to \infty$ the full dynamical mean-field theory is recovered which is thereby
recognized as the prototypical non-perturbative conserving approximation.
Clearly, in the case of the DMFT, $\Phi$-derivability is well known \cite{GKKR96}
and obvious, for example, when constructing the DMFT with the help of the 
skeleton-diagram expansion.

Using as a trial self-energy the self-energy of a cluster with $L>1$ correlated sites, 
generates an approximation where short-range spatial correlations are included up to the
cluster extension.
These variational cluster approximations provide a first step beyond the mean-field concept. 
Again, whether or not the sum rule is respected within the VCA depends on the reference
system itself.
For the $L=2$ Hubbard cluster, analytical calculations straightforwardly show that violations 
of the sum rule occur at half-filling, beyond a certain critical interaction strength.
In the thermodynamic limit, this would correspond to the Mott-insulating regime.
Applying the Lanczos method to larger clusters, has shown, however, that a breakdown of the
Luttinger sum rule is also possible for fillings off half-filling. 
For sufficiently strong $U$, the sum rule is violated in the whole $N=L-1$-particle
sector.
This would correspond to a (strongly correlated) metallic state in the thermodynamic limit.
Whether or not a VCA calculation is consistent with the sum rule, then depends on the
set of cluster hopping parameters $\ff t'$ which make the self-energy functional stationary.
First VCA calculations \cite{BHP07} for the $D=2$ Hubbard model at low doping and using 
clusters with up to $L=10$ sites do predict a violation in fact. 

It is by no means clear {\em a priori} what happens in a cluster approach using additional
bath degrees of freedom as variational parameters, as e.g.\ in the cellular DMFT.
\cite{KSPB01,LK00}
The usual periodization of the self-consistent C-DMFT self-energy, however, should be avoided
when testing the sum rule as this introduces an additional (though physically motivated)
approximation.
Instead, Eq.\ (\ref{eq:lutt1}) must be used with $\ff k$ re-interpreted as an index referring 
to the elements of the self-consistent diagonalized lattice Green's function. 

Employing the dynamical cluster approximation (DCA) \cite{HTZ+98}
represents an alternative which directly operates in reciprocal space.
From a real-space perspective, the DCA is equivalent with the cellular DMFT but applied to 
a modified model $H = H(\ff t, \ff U) \to H(\overline{\ff t}, \ff U)$ with modified hopping
parameters which are invariant under superlattice translations as well as under translations
on the cluster. \cite{BPK04,PB07}
In the limit $L\to \infty$ the replacement $\ff t \to \overline{\ff t}$ becomes irrelevant.
Analogous to the C-DMFT, the sum rule then holds within the DCA if and only if it holds for 
the individual cluster at self-consistently determined cluster parameters.
Note, however, that this requires that (besides the DCA self-energy) the {\em modified} 
hopping $\overline{\ff t}$ instead of the physical hopping has to be considered in the 
computation 
of the volume enclosed by the Fermi (Luttinger) surface of the lattice model.
This is exactly what is usually done in DCA calculations. 

Within this context and in view of the violations found for finite Hubbard clusters,
it is possible to understand why a non-perturbative cluster approximation, like the VCA, 
\cite{BHP07} or a cluster extension of the DMFT, like the DCA, \cite{MPJ02} can produce 
results that are inconsistent with Luttinger's theorem. 

\acknowledgments
We thank Robert Eder and Achim Rosch for valuable discussions. 
The work is supported by the Deutsche Forschungsgemeinschaft within the 
Forschergruppe FOR 538.

\appendix

\section{Macroscopic conservation of energy, particle number and spin}
\label{sec:cons}

The one-particle Green's function as obtained within an approximation generated 
by the choice of a reference system respects the macroscopic conservation laws 
which result from symmetries of the system with respect to continuous transformation
groups:

Energy conservation is apparently respected as {\em by construction} the approximate
SFT Green's function depends on a single frequency only, i.e.\ is invariant under
time translations. 

Conservation of the total particle number and spin is respected if the approximate
$\ff G$ transforms in the same way as the exact Green's function under global 
U(1) and SU(2) gauge transformations. 
Consider a general transformation of the form
\begin{equation}
  c^\dagger_\alpha \to \overline{c}^\dagger_\alpha 
  = \sum_\beta S_{\beta \alpha} c^\dagger_\beta \; 
\label{eq:str}  
\end{equation}
with unitary $\ff S$ such that the interaction part $H_1(\ff U)$ of the Hamiltonian
is invariant ($\alpha$ refers to the states of the one-particle basis).
In a diagrammatic approach, the invariance of $H_1(\ff U)$ implies that the corresponding 
conservation law is respected ``locally'' at each vertex.
Hence, for a conserving approximation in the sense of Baym and Kadanoff, the transformation 
behavior of the free Green's function is then propagated by the diagram rules to the full 
Green's function. 
Consequently, the latter must transform under $\ff S$ in the same way as the exact $\ff G$, i.e.
\begin{equation}
  G_{\alpha\beta} 
  \to \overline{G}_{\alpha\beta} = \left( \ff S \ff G \ff S^\dagger \right)_{\alpha\beta} \: .
\label{eq:trans}  
\end{equation}

Consider now the case of the SFT.
One has to show that the approximate Green's function $\overline{\ff G}$ for the 
transformed system with Hamiltonian $\overline{H}$ is given by 
$\overline{\ff G} = \ff S \ff G \ff S^\dagger$ if $\ff G$ is the approximate Green's 
function of the model $H$.
Applying the transformation (\ref{eq:str}) to $H$, one finds
$H = H_0(\ff t) + H_1(\ff U) \to \overline{H} = H_0(\overline{\ff t}) + H_1(\ff U)$ 
with $\overline{\ff t}=\ff S \ff t \ff S^\dagger$.
Again, $\ff S$ is assumed to leave the interaction part invariant.

The Green's function $\overline{\ff G}$ of the transformed model is (approximately) 
constructed via
\begin{equation}
   \overline{\ff G} = \frac{1}{\ff G^{-1}_{\overline{\ff t},0} 
   - \ff \Sigma_{\overline{\ff t}'_{\rm s},\ff U}}
\label{eq:dy1}
\end{equation}
from the free Green's function of the transformed model and the SFT self-energy
which is the self-energy of the reference system
$H'=H_0(\overline{\ff t}'_{\rm s})+H_1(\ff U)$
at the stationary point $\overline{\ff t}'_{\rm s}$.

For the transformed problem $\overline{H}$, the stationary point $\overline{\ff t}'_{\rm s}$ 
is determined from the SFT Euler equation:
\begin{equation}
   \sum_{\omega\alpha \beta} \left(
   \frac{1}{\ff G^{-1}_{\overline{\ff t},0} 
   - \ff \Sigma_{\overline{\ff t}',\ff U}}
   - \ff G_{\overline{\ff t}',\ff U}
   \right)_{\omega; \beta\alpha}
   \frac{\partial (\ff \Sigma_{\overline{\ff t}',\ff U})_{\omega; \alpha\beta}}
   {\partial \ff t'} = 0 \: .
\label{eq:eu1}   
\end{equation}
As an {\em ansatz} to solve the Euler equation we take 
\begin{equation}
\overline{\ff t}' = \ff S \ff t'_1 \ff S^\dagger
\end{equation} 
with $\ff t'_1$ to be determined.
The transformation law (\ref{eq:trans}) for the {\em exact} Green's function of
the reference system is
${\ff G}_{\overline{\ff t}',\ff U} = 
{\ff G}_{\ff S \ff t_1' \ff S^\dagger,\ff U} = 
\ff S \ff G_{\ff t'_1,\ff U} \ff S^\dagger$.
This also holds for the free Green's function.
Using the Dyson equation of the 
reference system we can deduce 
${\ff \Sigma}_{\overline{\ff t}',\ff U} = 
\ff S \ff \Sigma_{\ff t'_1,\ff U} \ff S^\dagger$.
Furthermore, for the free Green's function of the
transformed original model we have 
${\ff G}_{\overline{\ff t},0} = \ff S \ff G_{\ff t,0} \ff S^\dagger$.
Using these results, we see that Eq.\ (\ref{eq:eu1}) is equivalent to 
\begin{equation}
   \sum_{\omega\alpha \beta} \left(
   \frac{1}{\ff G^{-1}_{\ff t,0} 
   - \ff \Sigma_{\ff t'_1,\ff U}}
   - \ff G_{\ff t'_1,\ff U}
   \right)_{\omega; \beta\alpha}
   \frac{\partial (\ff \Sigma_{\ff t'_1,\ff U})_{\omega; \alpha\beta}}
   {\partial \ff t'} = 0 \: .
\label{eq:eu2}   
\end{equation}
But this is just the Euler equation for the original model which is solved
by $\ff t'_1 = \ff t'_{\rm s}$.
Remembering the ansatz made, we now have for the stationary point
$\overline{\ff t}'_{\rm s} = \ff S \ff t'_{\rm s} \ff S^\dagger$.
Inserting this into Eq.\ (\ref{eq:dy1}) gives
$\overline{\ff G} = \ff S (\ff G^{-1}_{\ff t,0} 
- \ff \Sigma_{\ff t'_{\rm s},\ff U})^{-1} \ff S^\dagger
= \ff S \ff G \ff S^\dagger$ which is the desired result.

Note that one has to ensure that the stationary point for the transformed problem 
$\overline{\ff t}'_{\rm s} = \ff S \ff t'_{\rm s} \ff S^\dagger$ lies within the 
space of one-particle parameters characteristic for the reference system. 
For models with local interaction part and for local (and also global) gauge 
transformations, however, this is always easily satisfied.

\end{document}